# Surface plasmon-mediated photoluminescence boost in graphene-covered CsPbBr$_3$ quantum dots

Youngsin Park[a,1], Elham Oleiki[a,1], Guanhua Ying[b], Atanu Jana[c], Mutibah Alanazi[b], Vitaly Osokin[b], Sangeun Cho[c], Robert A. Taylor[b,*], Geunsik Lee[a,*]

[a] Department of Chemistry, College of Natural Science, Ulsan National Institute of Science and Technology, Ulsan 44919, Korea

[b] Clarendon Laboratory, Department of Physics, University of Oxford, Parks Road, Oxford OX1 3PU, UK

[c] System Semiconductor Science, Dongguk University, Seoul 04620, Korea

ABSTRACT:

The optical properties of graphene (Gr)-covered CsPbBr$_3$ quantum dots (QDs) were investigated using micro-photoluminescence spectroscopy, revealing a remarkable three-orders-of-magnitude enhancement in photoluminescence (PL) intensity compared to bare CsPbBr$_3$ QDs. To elucidate the underlying mechanisms, we combined experimental techniques with density functional theory (DFT) calculations. DFT simulations showed that the graphene layer generates interfacial electrostatic potential barriers when in contact with the CsPbBr$_3$ surface, impeding carrier leakage from

[*] Corresponding author:

E-mail address: robert.taylor@physics.ox.ac.uk (R. Taylor), gslee@unist.ac.kr (G. Lee)

[1] These authors contributed equally to this work.

perovskite to graphene and enhancing radiative recombination. Additionally, graphene passivates $CsPbBr_3$ surface defect states, suppressing nonradiative recombination of photo-generated carriers. Our study also revealed that graphene becomes n-doped upon contact with $CsPbBr_3$ QDs, activating its plasmon mode. This mode resonantly couples with photo-generated excitons in the perovskite. The momentum mismatch between graphene plasmons and free-space photons is resolved through plasmon scattering at $Gr/CsPbBr_3$ interface corrugations, facilitating the observed super-bright emission. These findings highlight the critical role of graphene as a top contact in dramatically enhancing $CsPbBr_3$ QDs' PL. Our work advances the understanding of graphene-perovskite interfaces and opens new avenues for designing high-efficiency optoelectronic devices. The multifaceted enhancement mechanisms uncovered provide valuable insights for future research in nanophotonics and materials science, potentially leading to breakthroughs in light-emitting technologies.

1. **Introduction**

All inorganic lead halide perovskites (LHP) CsPbX$_3$ (X = Cl, Br or I) have emerged as promising materials for a wide range of optoelectronic applications such as solar cells [1,2], light-emitting diodes [3], and photodetectors [4] due to their higher stability compared to their organic-inorganic counterparts [5-7]. Among these, CsPbBr$_3$ has excellent optoelectronic properties, including an extremely high photoluminescence (PL) quantum yield, narrow emission bandwidth, and long carrier diffusion length and lifetime [8,9]. Furthermore, the emission energy can be tuned from 2.29 eV to 2.53 eV by synthesizing it in various structures such as bulk single crystals, thin films, nanocrystals (NCs), nanowires, and QDs [10-12], offering versatility in device design and application. CsPbBr$_3$ nanostructure emission could originate from various sources, including a single exciton, lasing due to biexcitonic emission [13], cavity lasing from a micro/nanowire [14], triplet excitonic emission [15], and superfluorescence [16]. Enhancing the optoelectronic properties of LHPs through heterojunction formation has been a focus of recent research. In particular, graphene (Gr), with its ultrahigh carrier mobility [17] and wavelength-independent light absorption [18], has been hybridized with CsPbBr$_3$ to improve its performance [4,19,20]. Previous studies have reported enhanced visible light absorption in CsPbBr$_3$-NCs/Gr heterostructures [19,20] and improved photoresponse due to facilitated electron transfer from CsPbBr$_3$ to graphene [4,19]. However, these configurations often led to PL quenching, limiting their potential in light-emitting applications. Surface defects in LHPs can introduce nonradiative trap states, significantly affecting their PL properties [21-23]. Various approaches have been explored to passivate these defects, including the application of different functional molecules [24]. Notably, encapsulating MAPbBr$_3$ NCs with Gr was shown to passivate surface defect states, enhancing PL intensity by a factor of seven [25]. Another effective method to improve the quantum efficiency of optoelectronic heterojunctions is PL enhancement through the resonant excitation of surface plasmons [26-28]. For Gr/semiconductor interfaces, it has been reported that a Gr plasmon can be resonantly activated by radiative recombination of electron-hole pairs from a semiconductor photo absorber, leading to enhanced PL through interaction with interface corrugations [29-32]. In this study, we present a novel approach to dramatically enhance the PL intensity of CsPbBr$_3$ QDs by covering them with Gr, achieving a three orders of magnitude increase. We employ a comprehensive set of experimental techniques, including μ-PL measurements and transmission electron microscopy (TEM), complemented by density functional theory (DFT) calculations to elucidate the underlying mechanisms. Crucially, we demonstrate that the configuration of Gr-covered CsPbBr$_3$ QDs (Gr/CsPbBr$_3$/SiO$_2$) leads to the formation of interfacial electrostatic potential barriers, which inhibit carrier leakage from the perovskite to the graphene layer. This is in stark contrast to the previously reported CsPbBr$_3$/Gr/SiO$_2$ configuration, where PL quenching was observed due to electron transfer from CsPbBr$_3$ to Gr [19].

## 2. Experimental section

2.1 Materials and Methods

In a 100 ml glass bottle, Cs$_2$CO$_3$ (0.0325 g, 0.01 mmol) was placed into a 15 mL glass vial with 2 mL 1-octadecene (ODE) and 1 mL oleic acid (OA), then dried at 120 °C for

1 hour until all Cs$_2$CO$_3$ had reacted with OA and kept it for 3 hr. In another glass vial, PbBr$_2$ (0.073 gm, 0.02 mmol) and n-octylammonium bromide (OAmBr) (0.042 gm, 0.2 mmol) were dissolved in 1 mL DMF and 2 mL of ODE at 120 ˚C and kept for 3 hr. Subsequently, 3 mL Cs-oleate solution was injected into the lead precursor solution, resulting in a change of solution color to greenish yellow. The crude solution was centrifuged for 5 minutes at 5000 rpm after cooling to room temperature. The residue was redispersed in toluene by sonication and again centrifuged at 5000 rpm for 5 min. The obtained product was washed with 2 ml of methyl acetate and subjected to another centrifugation at 10000 rpm for 10 minutes. Finally, the product was dried in a vacuum oven at 60 °C for 12 hours. The CsPbBr$_3$ QDs were put in a toluene solution and the solution was treated under sonication for 10 min. Then the solution was dispersed on a SiO2 (300 nm)/Si substrate. Then the graphene was transferred on the CsPbBr$_3$ dispersed SiO$_2$ substrate. The PMMA-coated Gr prepared on a Cu foil was placed in a solution of Cu etchant (CE-100, Transene Company, Inc.) to remove the Cu foil. After the Cu foil was completely removed away, the PMMA-coated Gr was scooped using the CsPbBr$_3$ dispersed SiO$_2$. The PMMA layer was then removed with acetone, and the sample was rinsed several times with deionized water.

2.2 Structural Characterization.

Transmission electron microscopy images were taken on a JEOL JEM-2100F electron microscope using a 200kV electron source. Samples were prepared on 200-mesh carbon coated Cu grids by dropping NC solutions which were allowed to evaporate.

## 2.3 Laser system and optical photoluminescence.

A 100 fs frequency-doubled Ti:sapphire laser operating at 400 nm with a repetition rate of 80 Mhz was used as the excitation source. The incident laser power on the $CsPbBr_3$ surface ranged from a few nW to a few tens of µW. The sample was mounted in a continuous-flow helium cryostat, allowing the temperature to be controlled accurately from 4.2 K to room temperature. Measurements were performed at cryogenic temperatures to resolve any fine structure, since its energy is small compared to the thermal bath at room temperature. The $CsPbBr_3$ nanocrystals in a toluene solution were dispersed on an Au patterned $SiO_2$ substrate to determine the position of nanocrystals with different sizes. The optical properties were characterized using a fiber-based confocal micro-photoluminescence (µPL) setup. A 100× long working distance apochromatic objective was held by a sub-micron precision piezoelectric stage above the cryostat and used to focus the incident laser beam to a spot size of ~1 µm$^2$ and to collect the resulting luminescence. The luminescence was then directed to a 0.3 m focal length spectrometer with a 600 gr/mm grating. The signal was finally detected using a cooled charge coupled device (CCD) detector. A telecentric lens arrangement was also present allowing the incident angle of the exciting laser at the entrance to the objective to be varied by a computer-controlled mirror thus providing an independent means to move the exiting spot relative to the collected emission, which is imaged confocally through the center of the objective. Time resolved photoluminescence (TRPL) measurements were carried out using the same experimental set up as above. The dispersed PL was reflected towards a photomultiplier connected to a commercial photon counting system. Measurements of the lifetimes of the confined states were then carried out over a range of excitation power densities.

2.4 Density functional theory calculations

DFT calculations were performed using the Vienna *ab initio* simulation package based on the plane wave pseudopotential approach [33], and by using a Generalized gradient approximation (GGA) within PBE parameterization [34] as the exchange-correlation functional. To incorporate van der Waals interactions, dispersion correction was added to the total energy and forces by employing the Tkatchenko and Scheffler (TS) approach [35] To model the $CsPbBr_3$ layer, we first optimized the atomic structure of the bulk $CsPbBr_3$ cubic phase using the experimental lattice parameter of 5.87 Å [36]. Then we used the optimized bulk unit cell and designed four different (001) $CsPbBr_3$ slabs considering both CsBr and $PbBr_2$ surface terminations [35,37]. To have distinguishable contributions from surface and bulk states in the electronic structure, we repeated the unit cell 7 times along the [001] direction and optimized all atomic positions. Also, to gain the lowest amount of lattice mismatch in heterostructure supercells, we employed $2 \times \sqrt{2} \times \sqrt{2}$ unit cells of 2D-$CsPbBr_3$, $7 \times 2$ rectangular layer of Gr, and $4 \times 2$ unit cells of SiO2 composed of 4 atomic layers along [001] direction. We made the $SiO_2$ slab from its cubic bulk structure with the lattice parameter of 4.93 Å. The lattice mismatch values for $CsPbBr_3$, Gr, and SiO2 are 0.8%, 6%, and 30%, respectively. To avoid spurious interactions between the periodic images of heterostructures we inserted 15 Å of vacuum along [001] direction and optimized all the atomic positions except two bottom layers of $SiO_2$ by using a single Γ-point. The kinetic energy cutoff was set to 400 eV while convergence criteria of $10^{-4}$ eV and 0.05 eV/Å were employed for energies and forces, respectively.

## 3. Results and Discussion

The detailed crystal structure of the individual $CsPbBr_3$ QDs was investigated using TEM. Figure 1a shows TEM images of the $CsPbBr_3$ QDs, revealing that the QDs tend to aggregate, forming clusters of various sizes. In contrast to previous reports of well-aligned $CsPbBr_3$ QDs [8,38-44], our QDs exhibit a random distribution ranging from approximately 10 to 50 nm, as shown in high-resolution TEM images (Fig. 1b). Despite this random size distribution, 60–70% of the QDs within each cluster are aligned in the same direction. When we measured the PL, we dispersed the QD clusters with a size of ~2 μm on the $SiO_2$ substrates. The schematic diagram of the micro PL system is depicted (Fig. 1c). The PL from these QD clusters may exhibit complex optical properties. While individual QDs in the cluster might emit independently, leading to a reduction in overall coherence, the close proximity of QDs within each cluster could potentially lead to collective effects. These effects might include exciton coupling or energy transfer between neighboring QDs, which could influence the emission characteristics. Furthermore, the orientation of QDs within a cluster, if sufficiently aligned, might contribute to some degree of directional emission. However, given the random distribution of QD sizes and orientations observed in the TEM images, the exact nature of these optical properties requires further investigation through detailed spectroscopic measurements and theoretical modeling. The resulting PL emission may exhibit complex characteristics, potentially including some degree of coherence and directionality, depending on the specific arrangement and interactions within each cluster. Like other semiconductor QDs, the size and geometry for a matrix of $CsPbBr_3$ QDs are subject to some extent of variations. Although the spin coating process has helped to spatially filter the QD clusters based on their sizes, we often found that clusters of similar dimensions are clumped together. It is therefore possible to excite multiple QD arrays under slightly different conditions within a single laser spot. In most cases, the QD clusters produce independent signals given their distinct emission centers. However, clusters with very similar geometric formations or even two closely alike matrices of QDs within a single cluster may be in close proximity, causing their individual collective behaviors to overlap and interact with each other.

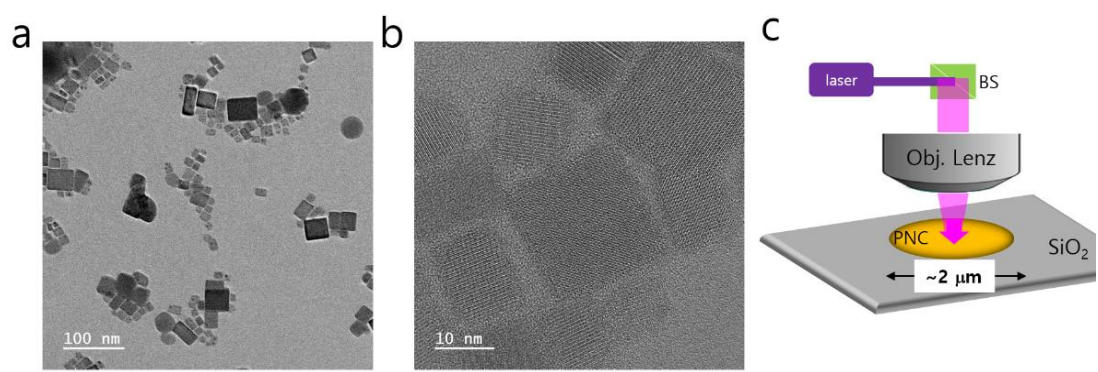

**Fig. 1.** (a) TEM image of $CsPbBr_3$ QD clusters, showing aggregation of individual QDs into various sized clusters. Scale bar: 100 nm. (b) High-resolution TEM image of individual QDs within a cluster, revealing the crystal structure and size distribution ranging from ~10 to 50 nm. Scale bar: 10 nm. (c) Schematic diagram of the µPL setup. The laser beam is focused on a $CsPbBr_3$ QD cluster with a diameter of ~2 µm on a $SiO_2$ substrate. BS represents the beam splitter, and Obj. indicates the objective used to focus the laser and collect the emitted light.

Figure 2 presents excitation power-dependent PL spectra for bare and Gr-covered CsPbBr$_3$ QDs, measured at low temperature. For bare CsPbBr$_3$ QDs (Fig. 2a), a single excitonic emission peak at 2.33 eV with a full width at half maximum (FWHM) of ~8 meV is observed at the lowest excitation fluence of 31.8 μJ/cm². As excitation fluence increases, new blue-shifted emission lines emerge, and the integrated intensity shows an S-shaped behavior (Fig. S1a), indicating stimulated emission. In contrast, Gr-covered CsPbBr$_3$ QDs (Fig. 2b) exhibit multiple emission peaks even at a low excitation fluence of 6 nJ/cm², with significantly enhanced PL intensity and signal-to-noise ratio. No new peaks emerge or blue-shift below excitation fluences of a few tens of μJ/cm². The integrated PL intensity increases linearly with excitation fluence (Fig. S1b). On the other hand, the PL intensity of the CsPbBr$_3$/Gr/SiO$_2$ heterostructure is quenched (Fig. 2c), consistent with the previous reports [4,19,20]. Even at excitation power of 0.16 kJ/cm$^2$, the spectrum is very noisy and broad with an FWHM of ~5 meV and the intensity doesn't increase much at high excitation power of 2.7 kJ/cm$^2$. Figure 2d compares PL spectra for the three different heterostructures, demonstrating significant PL enhancement in the Gr-covered structure despite the absence of stimulated emission peaks. To investigate the origin of the PL intensity quenching, we explored the electronic structure of the CsPbBr$_3$/Gr heterostructure by performing the DFT calculations.

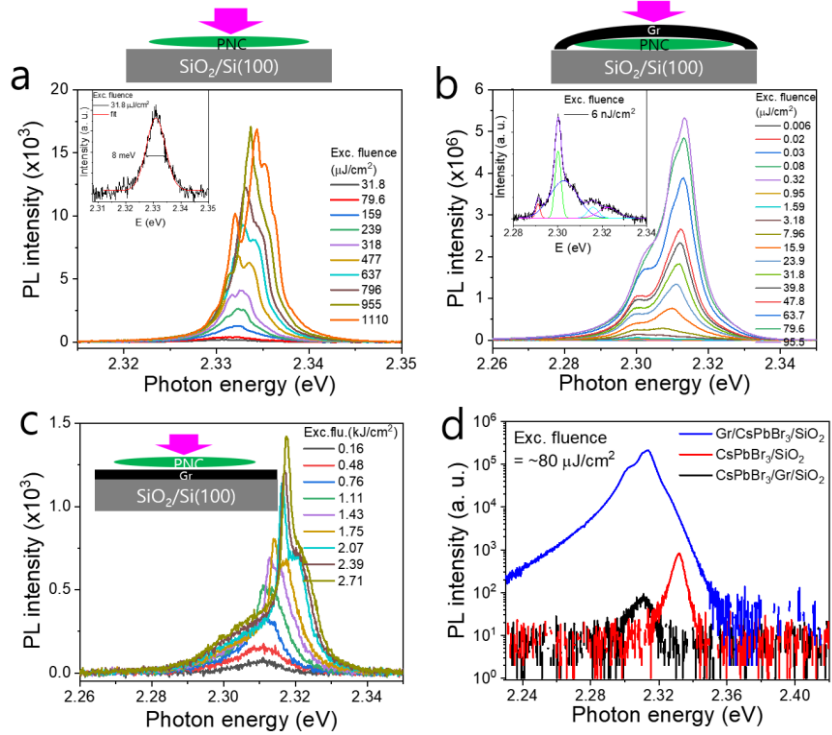

**Fig. 2.** (a) Power-dependent μPL spectra measured for the bare CsPbBr$_3$ QD cluster. Inset depicts the PL spectrum and fit (red) at the lowest excitation fluence. (b) Power-dependent μPL spectra taken for a Gr-coated CsPbBr$_3$ heterostructure cluster. Inset depicts the PL spectrum and fit at the lowest excitation fluence. (c) The power-dependent μPL spectra of the CsPbBr$_3$/Gr structure. (d) Comparison of PL spectra for Gr/CsPbBr$_3$ and CsPbBr$_3$ structures at similar excitation fluence (~80 μJ/cm²). For the CsPbBr$_3$/Gr, the excitation fluence is ~160 μJ/cm². Schematics above each graph illustrate the corresponding sample structure.

The model structure is described in the method section and is illustrated (Fig. 3a and Fig. S2). According to the DFT calculation results, the CsPbBr$_3$/Gr interface is stabilized through weak van der Waals interaction with an interlayer distance of 3.4 Å and binding energy of -0.03 eV per carbon atom (Eq. S1 in supporting information). The ground state electronic band structure of the CsPbBr$_3$/Gr/SiO$_2$ heterostructure, including a perovskite symmetry slab with PbBr$_2$ termination, is depicted in Fig. 3a.

The CsPbBr$_3$ layer shows a direct band gap of 1.9 eV located at the Γ point. In addition, upon forming CsPbBr$_3$/Gr interface, electrons transfer from the CsPbBr$_3$ to the graphene, indicating the n-doped character of graphene as evidenced by the Dirac point appearing below the Fermi level. Moreover, the Fermi level is strongly pinned to the perovskite's valence band maximum (VBM). Therefore, under the illumination of the perovskite light absorber, electrons will readily transfer from the graphene layer to photo-generated empty states in the perovskite valance band. Consequently, the radiative recombination of photo-excited electron-hole pairs in perovskite will be suppressed, leading to PL quenching in the CsPbBr$_3$/Gr/SiO$_2$ system, which agrees with previous reports [4,19,20]. The Fermi level pinning at the perovskite VBM is confirmed to appear when the PbBr$_2$-terminated surface is in contact with Gr/SiO$_2$, being independent of the terminal type of the opposite side (Figs. S3a-c).

We also modeled the Gr/CsPbBr$_3$/SiO$_2$ heterostructure (Fig. 3b and Fig. S2) to determine the reason for the PL intensity enhancement. Like the CsPbBr$_3$/Gr/SiO$_2$ heterostructure, the interfacial interaction at the Gr/CsPbBr$_3$ junction is of van der Waals type with an interlayer distance of 3.3 Å and a binding energy of -0.03 eV per carbon atom (Eq. S2 in supporting information). The ground state electronic band structure of the Gr/CsPbBr$_3$/SiO$_2$ heterostructure is shown in Fig. 3b. The CsPbBr$_3$ layer exhibits a direct band gap of 1.80 eV at the Γ point, which is slightly reduced compared with CsPbBr$_3$/Gr/SiO$_3$ heterostructure, and Gr's Dirac cone is located between the Γ and X points. Interacting with the SiO$_2$ interface dipole increases the CsPbBr$_3$ work function, leading to a downward shift of its VBM and CBM.

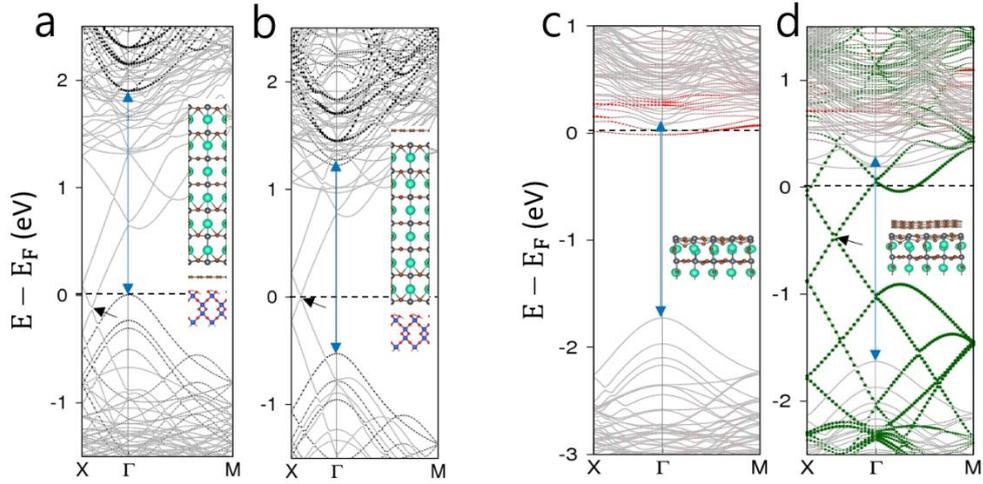

**Fig. 3.** Calculated band structure of (a) $CsPbBr_3$/Gr/$SiO_2$ (inset: side view along [100] of the model heterostructure). (b) Gr/$CsPbBr_3$/$SiO_2$. For easier recognition of the $CsPbBr_3$ band edges, the Pb bulk atom's 6p orbital contribution is shown by black circles. (c) $CsPbBr_3$ with a $V_{Br}$ surface defect (inset: side view along [100] of the top 4 atomic layers of the model structure). (d) a defected $CsPbBr_3$/Gr interface (inset: side view along the [100] interface model structure with a perovskite top 4 atomic layers). Red circles show the contribution of the under-coordinated Pb atoms, and the carbon $p_z$ orbital contribution is indicated by green circles. In all band structures, the electronic transition between the $CsPbBr_3$ band edges is indicated by the blue arrow. Gr's Dirac point is indicated by the black arrow in (a), (b), and (d). Atomic color scheme: Cs (green), Pb (gray), Br (brown), Si (blue), O (red), H (pink), C (brown).

This, in turn, enhances the interfacial electrostatic potential barrier for charge transfer at the Gr/$CsPbBr_3$ interface which is evident by comparing the position of the perovskite's VBM and CBM relative to the graphene Dirac point (Fig. 3b). Furthermore, following the downshift of the perovskite energy levels, the interfacial orbital coupling at the Gr/$CsPbBr_3$ interface is suppressed. Therefore, the formation of the

Gr/CsPbBr$_3$/SiO$_2$ heterostructure does not result in charge transfer at the Gr/perovskite interface in its ground state, even for the other termination types of the perovskite layer (Figs. S3d-f). Under illumination, the interfacial electrostatic potential barriers are high enough to prevent carrier leakage from the perovskite layer to graphene, which enhances radiative recombination compared to CsPbBr$_3$/Gr/SiO$_2$ heterostructure.

Another possibility for explaining the PL enhancement could be associated with a surface stabilization effect via defect passivation, as we have shown in our previous work on passivation effect giving rise to significant PL enhancement [25]. We explored the structural and electronic properties of a defected CsPbBr$_3$ layer with and without graphene on its surface. Since a Br vacancy (V$_{Br}$) is reported to be the predominant defect in CsPbBr$_3$ perovskites [23,45], we introduced a V$_{Br}$ into a PbBr$_2$-terminated CsPbBr$_3$ (001) surface (Fig. S4). The optimized geometry for the defected interface (Inset in Fig. 3d) indicates structural distortions in the perovskite top layers caused by V$_{Br}$. Gr is physisorbed on the perovskite surface with a binding energy of -0.03 eV per carbon atom (Eq. S3 in supporting information). We should note that the graphene deformation is an artifact of our model supercell. As shown in Fig. 3c, the V$_{Br}$ generates a localized trap state close to the CsPbBr$_3$ conduction band minimum (CBM), which can act as a center for nonradiative recombination and suppress the PL intensity. However, after covering defected CsPbBr$_3$ with graphene, carbon $p_z$ states hybridized with Pb dangling bond states and shifted them upward to the perovskite conduction band (Fig. 3d). Therefore, the Gr adsorption on the defected CsPbBr$_3$ surface eliminates nonradiative recombination centers leading to PL enhancement in the Gr/CsPbBr$_3$/SiO$_2$ system. It should be noted that upon passivation of the $V_{Br}$ defect, the Gr is n-doped (Fig. 3d) with a charge density of $10^{13}$ cm$^{-2}$ (Eq. S4 in supporting information) [46]. Similar results are obtained for the Gr/CsPbBr$_3$ interface consisting of the perovskite with a CsBr termination (Figs. S4 and S5).

Enhanced PL emission due to the resonant coupling between surface plasmon (SP) modes in a metallic layer and excitons in a semiconductor has been reported [31,42,47-51]. Considering the semi-metallic nature of graphene, the significant PL enhancement of the Gr/CsPbBr$_3$/SiO$_2$ heterosystem could be understood in terms of the resonant excitation of graphene plasmon modes [29-31,50]. The perovskite's photo-generated excitons can induce a resonant plasmon mode in the graphene layer. However, because of the considerable momentum mismatch between the Gr's plasmon and photons in free space, a lateral modulation periodicity on the graphene [29,30] or its substrate [31] is required to resolve the momentum mismatch and transform the excited plasmon mode to the light emission. The dispersion relation of the graphene plasmon in the random phase approximation is

$$\omega(q) = \left[\frac{n_e e^2}{\epsilon_0 (1+\epsilon_b) m^*} |q| + \frac{3}{4} v_F^2 q^2\right]^{1/2} \quad (1)$$

where $q$ is the in-plane wave number, $n_e$ is the electron density, $\epsilon_0$ is the vacuum permittivity, $\epsilon_b$ is the substrate static dielectric constant, $m^*$ is electron effective mass in Gr, and $v_F$ is its Fermi velocity [31,51,52]. We supposed that the graphene plasmon could be resonantly excited at perovskite band gap energy of 2.3 eV and calculated the in-plane momentum $q$ from our calculated $n_e = 10^{13}$ cm$^{-2}$, using $\epsilon_b = 18.6$ for the CsPbBr$_3$ substrate [21,53], $m^* = 0.077\ m_e$ where $m_e$ is the free electron mass, and $v_F =$

$1.12 \times 10^6$ m/s [31]. We obtained $q = 2\pi/1.8$ nm$^{-1}$, indicating a lateral modulation periodicity of $a = 1.8$ nm at the Gr/CsPbBr$_3$ interface would be required to extract light from the Gr's plasmon. Due to the V$_{Br}$ defects, the surface structure of the CsPbBr$_3$ is distorted and not perfectly flat, resulting in interface corrugations on the order of a few nm. While it may not be feasible to measure $a = 1.8$ nm experimentally, this value is consistent with the lateral distance of 1.68 nm between V$_{Br}$ surface defects in our model structure. Thus, the resonant activation of a Gr plasmon at the perovskite band gap, and its conversion to photons via interaction with interface surface corrugations, explains the observed PL enhancement for the Gr/CsPbBr$_3$/SiO$_2$ heterostructure. A schematic diagram of the explained mechanism is depicted in Fig. 4a. We also conducted TRPL measurements on bare and Gr-covered CsPbBr$_3$ NCs to understand the dynamics of photoexcited carriers. The PL signals of both samples show a biexponential decay over time as presented in Fig. 4b. The fast decay component of bare CsPbBr$_3$ NCs is 0.61 ns, which reduced to 0.5 ns for Gr-covered CsPbBr$_3$ NCs, originating from the direct recombination of photoexcited charge carriers [54,55] in both samples. On the other hand, the slow decay component indicates the presence of a secondary source for PL emission, attributed to shallow trap-assisted recombination (V$_{Br}$ in our case) [54]. The slow decay component for bare CsPbBr$_3$ NCs is 3.26 ns which reduces to 2.96 ns for the Gr-covered sample, confirming the decrease of surface defect states by Gr covering.

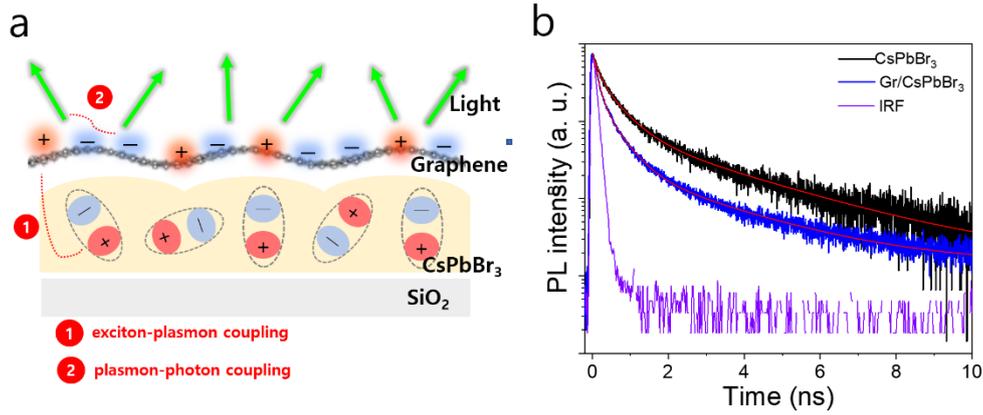

**Fig. 4.** (a) Schematic diagram of the emission from exciton-plasmon and plasmon-photon coupling. (b) Time-resolved PL comparison for bare CsPbBr$_3$ QDs (uncoated) and Gr-covered QD clusters. The data of bare CsPbBr$_3$ and Gr-covered CsPbBr$_3$ are fitted with biexponential decays and the decay times are 0.61 ns and 0.5 ns for faster components and 2.96 ns and 3.26 ns for slow components, respectively. The TRPL measurement accuracy is 120 ps.

## 4. Conclusions

In summary, we have demonstrated the optical characterization of bare CsPbBr$_3$ QDs and graphene-covered CsPbBr$_3$ QDs by micro-photoluminescence measurements and density functional theory calculations. Compared with bare CsPbBr$_3$ QDs, the PL intensity of the Gr-covered CsPbBr$_3$ QDs is dramatically enhanced by three orders of magnitude. The DFT calculation results showed that contacting Gr on the top side of CsPbBr$_3$ causes the Dirac point to appear at the center of CsPbBr$_3$ energy gap, in contrast to the Fermi level pinning behavior to the valence band edge reported for bottom side contact as in CsPbBr$_3$/Gr/SiO$_2$. Thus, carrier leakage is prevented for the Gr-covered CsPbBr$_3$ with a high radiative recombination rate. In addition, the perovskite's surface defects are passivated via the graphene covering, suppressing the nonradiative recombination of photo-generated charge carriers compared to uncovered perovskite QDs. Furthermore, resonant excitation of graphene plasmon by the perovskite's photo-generated excitons, followed by its conversion to photons through interaction with $Gr/CsPbBr_3$ interface corrugations, substantially enhances the photoluminescence from $Gr/CsPbBr_3/SiO_2$ heterosystem. Our experimental and theoretical investigations show the significant role of graphene covering for dramatic PL enhancement of $CsPbBr_3$, which can be implemented for developing super bright light-emitting diodes.

# Surface plasmon-mediated photoluminescence boost in graphene-covered CsPbBr$_3$ quantum dots

Youngsin Park [a], Elham Oleiki [a], Guanhua Ying [b], Atanu Jana [c], Mutibah Alanazi [b], Vitaly Osokin [b], Sangeun Cho [c], Robert A. Taylor [b,*], Geunsik Lee [a,*]

[a] Department of Chemistry, College of Natural Science, Ulsan National Institute of Science and Technology, Ulsan 44919, Korea

[b] Clarendon Laboratory, Department of Physics, University of Oxford, Parks Road, Oxford OX1 3PU, UK

c System Semiconductor Science, Dongguk University, Seoul 04620, Korea

[†]These authors are contributed equally to this work

*Correspondence authors. E-mail: robert.taylor@physics.ox.ac.uk, gslee@unist.ac.kr

**Density functional theory calculations**

To model the CsPbBr$_3$ surface with V$_{Br}$ defects, we used the ideal CsPbBr$_3$/Gr supercell, and repeated it twice along the [010] direction (Figure S4), and introduced one surface Br vacancy per each 1×2 supercell. Our defective model of CsPbBr$_3$ slab corresponds to the surface defect concentration of $10^{13}$ cm$^{-2}$, comparable to the experimental value [1,2]. Atomic positions of the Gr layer and four top atomic layers of the CsPbBr$_3$ surface were optimized. We performed PBE electronic structure calculations on a 4×4×1 *k*-point mesh by setting energy convergence criteria and kinetic energy cutoff $10^{-4}$ and 400 eV, respectively. Our calculated direct band gap for cubic bulk structure is 1.53 eV which is consistent with other theoretical reports [3]; however, it underestimates the experimental value of 2.22 eV [4]. To get a better agreement with the experimental band gap, GW+SOC or HSE+SOC methods should be implemented [3,5]; though, considering the high number of atoms in our supercells (more than 300), employing this level of calculation was not plausible. Nonetheless, the main characteristics of the conduction and valance bands of CsPbBr$_3$ are maintained using the PBE method [6]. In addition, because of the quantum confinement effect, the band gap of the CsPbBr$_3$ layer increased to 1.85 eV which is closer to the experimental value. Therefore, we performed all the electronic structure calculations using the PBE method.

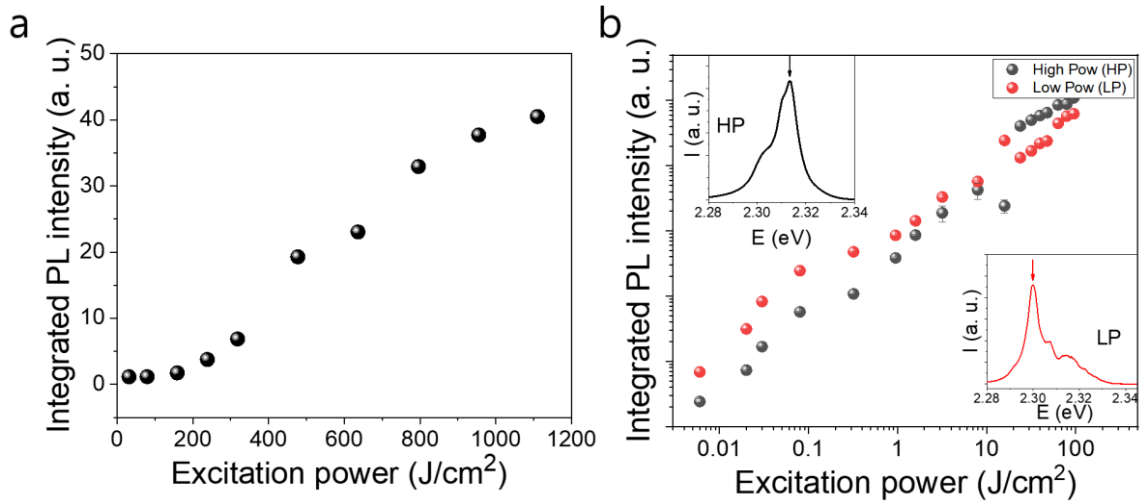

**Fig. S1**. (a) Integrated PL intensity of the bare CsPbBr$_3$ QDs red dot marked peak in Fig. 2a. (b) Integrated PL intensity of the Gr-covered CsPbBr$_3$ QDs at the main peaks of the low and high excitation power. Here, LP and HP mean the low excitation power and high excitation power, respectively.

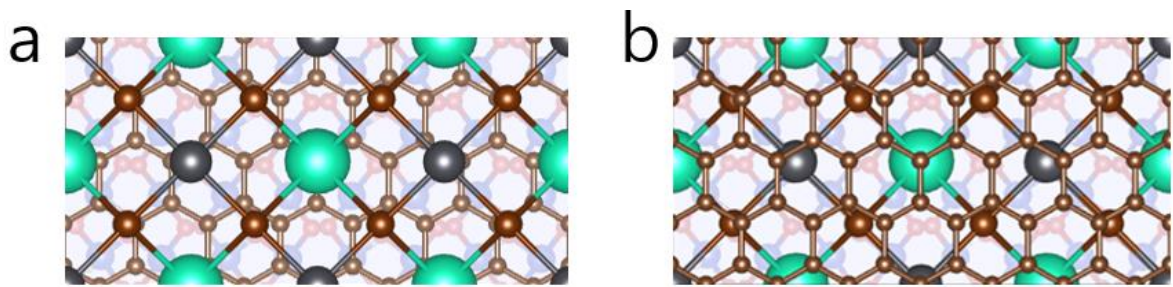

**Fig. S2**. Top view of $2 \times \sqrt{2} \times \sqrt{2}$ surface supercell of (a) CsPbBr$_3$/Gr/SiO$_2$, and (b) Gr/CsPbBr$_3$/SiO$_2$. Atomic color scheme: Cs (green), Pb (gray), Br (brown), Si (blue), O (red), H (pink), C (brown).

**Electronic band structure of the three-layered heterostructure considering CsPbBr$_3$ layer with CsBr and PbBr$_2$ terminations**

Considering both CsBr and PbBr$_2$ surface terminations of the perovskite layer, we calculated the ground state electronic band structure of four CsPbBr$_3$/Gr/SiO$_2$ (Figures S3a-c　Figure 3a) and four Gr/CsPbBr$_3$/SiO$_2$ model heterostructures (Figures 3d-e and Figure 3b) to screen all possible interfacial interactions and their impact on the electronic structure. Among all CsPbBr$_3$/Gr/SiO$_2$ model heterostructures, the model including a CsPbBr$_3$ (001) slab with PbBr$_2$ top and bottom terminations (Figure 3a) showed considerable charge transfer at the Gr/CsPbBr$_3$ interface. However, none of the Gr/CsPbBr$_3$/SiO$_2$ model heterostructures showed interfacial charge transfer at the CsPbBr$_3$/Gr interface.

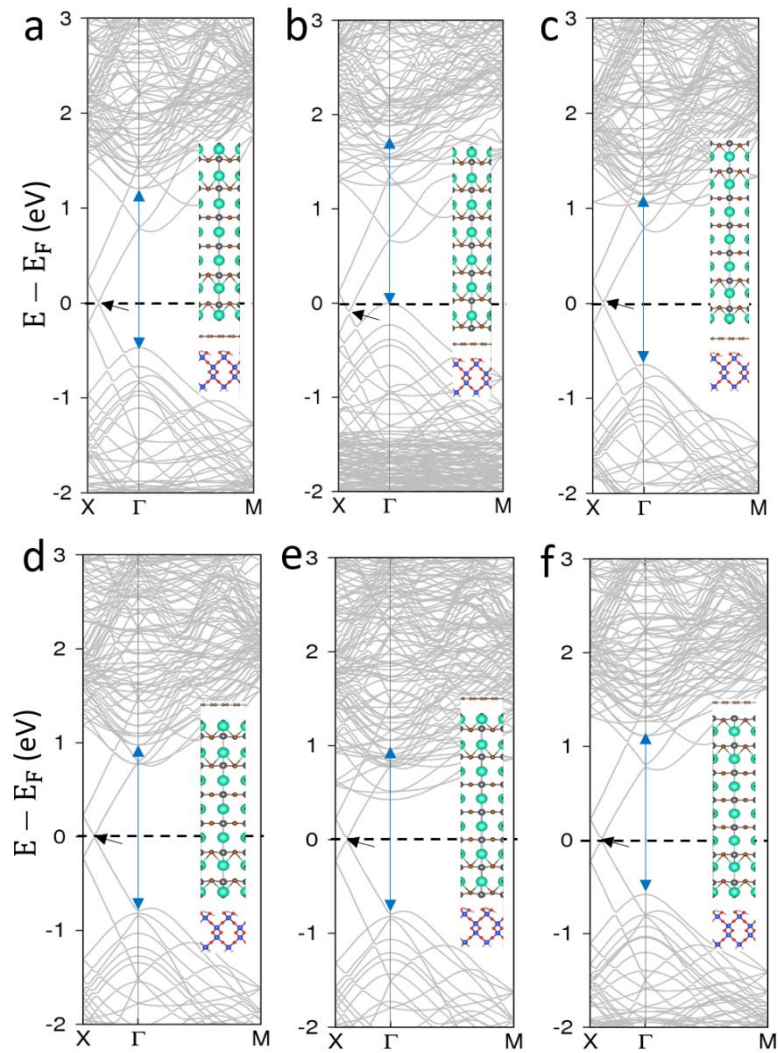

**Fig. S3.** Calculated electronic band structure of (a)(b)(c) CsPbBr$_3$/Gr/SiO$_2$, and (d)(e)(f) Gr/CsPbBr$_3$/SiO$_2$ with different perovskite surface terminations. The electronic transition between the CsPbBr$_3$ band edges is indicated by the blue arrow, and graphene's Dirac point is mentioned by the black arrow. All side views in the insets are along [100]. Atomic color scheme: Cs (green), Pb (gray), Br (brown), Si (blue), O (red), H (pink), C (brown).

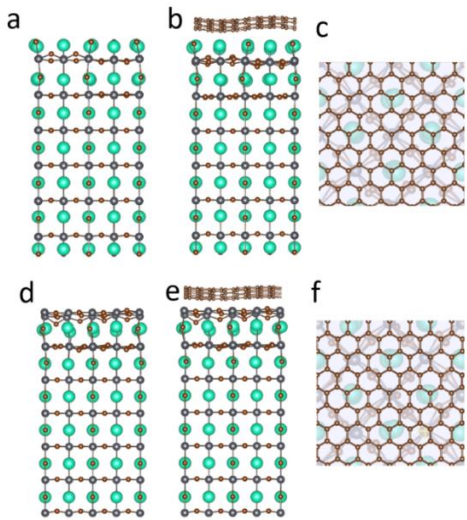

**Fig. S4**. Side view along [100] of (a) CsBr terminated (001) CsPbBr$_3$ slab with V$_{Br}$ surface defect and (b) its interface with Gr. (c) Top view of (b). (d) PbBr$_2$ terminated (001) CsPbBr$_3$ slab with V$_{Br}$ surface defect and (e) its interface with Gr. (f) Top view of (e). Atomic color scheme: Cs (green), Pb (gray), Br (brown), Si (blue), O (red), H (pink), C (brown).

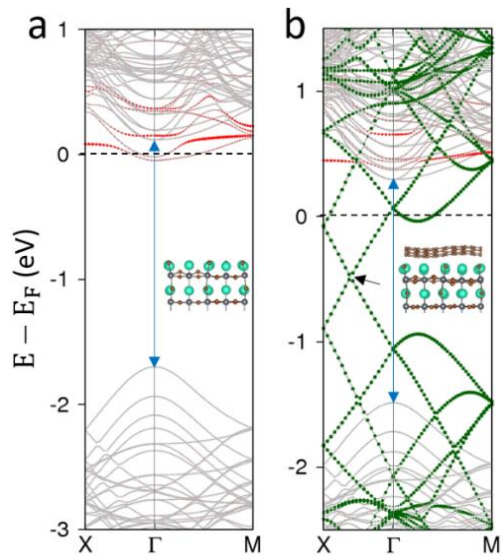

**Fig. S5.** Calculated electronic band structure of (a) CsBr terminated (001) CsPbBr$_3$ slab with V$_{Br}$ surface defect (inset: side view along [100] of the top 4 atomic layers of model structure). (b) defected CsPbBr$_3$/Gr interface (inset: side view along [100] of interface model structure with perovskite top 4 atomic layers). Red circles show the contribution of under-coordinated Pb atom, and carbon $p_z$ orbital contribution is indicated by green circles. In all band structures, CsPbBr$_3$ band edges are indicated by arrows. Atomic color scheme: Cs (green), Pb (gray), Br (brown), Si (blue), O (red), H (pink), C (brown).

**Interlayer binding energy**

Equations S1 to S3 define the binding energy of Gr and perovskite layer in $CsPbBr_3/Gr/SiO_2$, $Gr/CsPbBr_3/SiO_2$, and $Gr/CsPbBr_3$ heterostructure, respectively:

$$E_B = E\,(CsPbBr_3/Gr/SiO_2) - E\,(Gr/SiO_2) - E\,(CsPbBr_3) \quad (Eq.\ S1)$$

$$E_B = E\,(Gr/CsPbBr_3/SiO_2) - E\,(Gr) - E\,(CsPbBr_3/SiO_2) \quad (Eq.\ S2)$$

$$E_B = E(Gr/CsPbBr_3) - E\,(CsPbBr_3) - E\,(Gr) \quad (Eq.\ S3)$$

where $E\,(CsPbBr_3/Gr/SiO_2)$ is the total energy of $CsPbBr_3/Gr/SiO_2$ heterostructure, $E(Gr/SiO_2)$ is the total energy of $Gr/SiO_2$ heterostructure, $E\,(CsPbBr_3)$ the total energy of isolated $CsPbBr_3$, $E\,(Gr/CsPbBr_3/SiO_2)$ is the total energy of $Gr/CsPbBr_3/SiO_2$ system, $E\,(Gr)$ is the total energy of isolated Gr layer, and $E(Gr/CsPbBr_3)$ is the total energy of $Gr/CsPbBr_3$ heterostructure. According to these definitions, negative (positive) value of $E_B$ suggests a stable (unstable) interface.

**Electron density in the graphene layer**

Equation S4 relates the Dirac energy in electronic band structure to electron density in the Gr layer:

$$n_e = \frac{E_D^2}{\pi\,(\hbar v_F)^2} \quad (Eq.\ S4)$$

where $v_F$ is Fermi velocity. For $E_D = 0.5$ eV (from electronic band structure of Gr/defected-$CsPbBr_3$ interface (Figure 3d)) and $v_F = 1.12 \times 10^6$ m/s, the calculated $n_e$ value is $10^{13}$ cm$^{-2}$.